\documentclass[10pt,conference]{IEEEtran}
\usepackage{cite}
\usepackage{amsmath,amssymb,amsfonts}
\usepackage{algorithmic}
\usepackage{graphicx}
\usepackage{textcomp}
\usepackage{xcolor}
\usepackage{pifont}
\usepackage{balance}
\usepackage{url}

\usepackage{float}
\usepackage[frozencache=true,cachedir=minted-cache]{minted}

\let\svthefootnote\thefootnote
\newcommand\blankfootnote[1]{%
  \let\thefootnote\relax\footnotetext{#1}%
  \let\thefootnote\svthefootnote%
}
\let\svfootnote\footnote
\renewcommand\footnote[2][?]{%
  \if\relax#1\relax%
    \blankfootnote{#2}%
  \else%
    \if?#1\svfootnote{#2}\else\svfootnote[#1]{#2}\fi%
  \fi
}

\definecolor{commentgray}{rgb}{0.2,0.2,0.2}
\definecolor{bggray}{rgb}{0.98,0.98,0.98}

\setminted{
  bgcolor=white,
  numberblanklines=false,
  frame=single,
  numbers=left,
  tabsize=2,
  xleftmargin=1.8em,
  xrightmargin=0.5em
}

\def\BibTeX{{\rm B\kern-.05em{\sc i\kern-.025em b}\kern-.08em
    T\kern-.1667em\lower.7ex\hbox{E}\kern-.125emX}}
\begin{document}

\title{MLTEing Models: Negotiating, Evaluating, and Documenting Model and System Qualities\\
}

\makeatletter 
\newcommand{\linebreakand}{%
  \end{@IEEEauthorhalign}
  \hfill\mbox{}\par
  \mbox{}\hfill\begin{@IEEEauthorhalign}
}
\makeatother

\newcommand\sA{\ding{192}}
\newcommand\sB{\ding{193}}
\newcommand\sC{\ding{194}}
\newcommand\sD{\ding{195}}
\newcommand\sE{\ding{196}}

\author{
\centering
\IEEEauthorblockN{Katherine R. Maffey*}
\IEEEauthorblockA{\textit{AI Integration Center} \\
\textit{U.S. Army}\\
Pittsburgh, PA \\
kmaffey@alumni.cmu.edu}
\and
\IEEEauthorblockN{Kyle Dotterrer*}
\IEEEauthorblockA{\textit{AI Integration Center} \\
\textit{U.S. Army}\\
Pittsburgh, PA \\
kdottere@alumni.cmu.edu}
\and
\IEEEauthorblockN{Jennifer Niemann}
\IEEEauthorblockA{\textit{AI Integration Center} \\
\textit{U.S. Army}\\
Pittsburgh, PA \\
jmnieman@alumni.cmu.edu}
\linebreakand
\IEEEauthorblockN{Iain Cruickshank}
\IEEEauthorblockA{\textit{Army Cyber Institute} \\
\textit{United States Military Academy}\\
West Point, NY \\
}
\and
\IEEEauthorblockN{Grace A. Lewis}
\IEEEauthorblockA{\textit{Software Engineering Institute} \\
\textit{Carnegie Mellon University}\\
Pittsburgh, PA \\
}
\and
\IEEEauthorblockN{Christian Kästner}
\IEEEauthorblockA{\textit{Software and Societal Systems Dep.} \\
\textit{Carnegie Mellon University}\\
Pittsburgh, PA}
}

\maketitle

\begin{abstract}

Many organizations seek to ensure that machine learning (ML) and artificial intelligence (AI) systems work as intended in production but currently do not have a cohesive methodology in place to do so. To fill this gap, we propose MLTE (Machine Learning Test and Evaluation, colloquially referred to as ``melt''), a framework and implementation to evaluate ML models and systems. The framework compiles state-of-the-art evaluation techniques into an organizational process for interdisciplinary teams, including model developers, software engineers, system owners, and other stakeholders.  MLTE tooling supports this process by providing a domain-specific language that teams can use to express model requirements, an infrastructure to define, generate, and collect ML evaluation metrics, and the means to communicate results.
\end{abstract}

\begin{IEEEkeywords}
machine learning, test and evaluation, machine learning evaluation, responsible AI
\end{IEEEkeywords}

\section{Introduction}

Testing is an essential component of product development; in most science and technology domains, consumers assume that a product or piece of equipment is tested for safety considerations before it becomes widely available for use. For instance, US law requires that pharmaceutical companies test drugs before making them available for consumption \cite{darrow2020fda}, and the Federal Aviation Administration verifies every aircraft's airworthiness \cite{de2016airworthiness, faa2022airworthy}. Despite this norm, there is a lack of accepted best practices for testing \textit{software products that contain machine learning components}, hereafter abbreviated as \textit{ML systems}. As ML gains significant attention in several domains and for many real-world problems, the lack of testing standards becomes more prominent. \blankfootnote{{*} Both authors contributed equally to this research.}

Testing ML systems continues to be challenging \cite{amodei2016concrete, hendrycks2021unsolved}. While any large software system is complex to develop and deploy \cite{zhang2003machine}, the introduction of ML components introduces additional challenges such as defining comprehensive requirements and evaluation criteria to create high-performing models \cite{studer2021towards}. A lack of rigorous ML evaluation can manifest in negative real-world outcomes. Amazon's gender-biased ML resume vetting tool and the fatal crash caused by a Tesla Model S operating under Autopilot in May 2016 are two examples that demonstrate this reality \cite{dastin2018amazon, banks2018driver}. As more organizations seek to develop, deploy, and operate ML systems in a growing number of contexts, it is clear that evaluating these systems is important and necessary despite presenting significant challenges~\cite{ashmore2021assuring, braiek2020testing, hendrycks2021unsolved, holstein2019improving, doshi2017towards}.

In this paper, we define \textit{Machine Learning Test and Evaluation} or \textit{MLTE} (pronounced ``melt''), a framework and implementation that guide organizations through the ML evaluation process. Based on state-of-the-art ML research, MLTE gives teams a tool to more effectively \textit{negotiate, evaluate, and document} results throughout ML system evaluation. While our focus is ML systems, MLTE is motivated by documented challenges from throughout the ML development life cycle that make defining and evaluating success difficult. Many issues that arise in the ML development process and manifest in production are the result of requirements engineering shortcomings \cite{ishikawa2019, villamizar2021requirements}; accordingly, our proposal emphasizes this element of the procedure. To effectively address as many challenges as possible, the MLTE process facilitates evaluation from the inception of a ML project and progresses with it through completion. While many existing ML workflows include evaluation as a step along the path to a finished ML system, we propose that ML system evaluation must be considered throughout the life cycle of a project.

\section{Open Problems and State of the Art}
\label{sec:open-problems}

Historically, ML research has focused on advancing model capability, typically in terms of accuracy achieved on benchmark problems \cite{wagstaff2012machine}. Despite the growing adoption of ML components in software products, most education and research still has a model-centric view that rarely considers the system beyond the model \cite{kaestner_model_to_system}. Many practical problems faced by engineering teams happen \textit{at the boundary} between ML components and other parts of the system: examples include unstable data dependencies and hidden feedback loops \cite{sculley2015hidden}, limited considerations of system-level qualities like fairness and safety \cite{holstein2019improving,borg2019}, and poor collaboration between data scientists and software engineers \cite{nahar2022collaboration}. In the following sections, we describe three open problems which are representative of commonly-reported issues throughout the ML development process.

\subsection{Problem 1: Communication Barriers}
\label{sec:comms-barriers}

Research into applications of ML systems and organizational processes used to implement them reveals that communication is a critical yet often neglected part of product development \cite{nahar2022collaboration, paleyes2022}. Organizations often silo team members into distinct software engineering and data science roles; intra-team isolation is exaggerated in organizations that outsource model development. Isolating team members creates communication barriers that often surface as problems during integration when team members realize that particular requirements were not communicated \cite{nahar2022collaboration, sculley2015hidden}. For example, data scientists may deliver a model that meets accuracy needs but has an inference latency that is unacceptably high for production deployment. Another critical but often ignored aspect of the development process is negotiating requirements for the system and individual components \cite{vogelsang2019requirements, nahar2022collaboration, studer2021towards, kaestner_model_to_system}. During model development, teams must consider requirements beyond accuracy such as inference costs, data quality, data quantity, robustness, or fairness requirements \cite{siebert2020towards,nahar2022collaboration, vogelsang2019requirements}. Failing to define and communicate requirements may lead to a mismatch of assumptions from involved entities, which can result in system failures \cite{lewis2021characterizing}. 

A final common communication barrier is the process of communicating ML evaluation results amongst teams and to external stakeholders \cite{rakova2021responsible}. Tools such as MLflow \cite{MLflow} and Weights and Biases \cite{weights} assist model developers with recording experimentation procedures but fail to capture the connection between individual metrics and system outcomes. The \textit{model cards} \cite{mitchell2019model} proposal provides an example of ML results reporting but is intended to document the model rather than the evaluation process it undergoes. Given the existing challenges in sharing evaluation results with users \cite{liu2020}, organizations need a method to effectively, consistently, and automatically communicate the results of ML model evaluation. 

To address these communication barriers, organizations require a process that establishes consistent communication points, facilitates requirements definition, and offers automated reporting.

\subsection{Problem 2: Low-Quality or Missing Model Documentation}
\label{sec:documentation}

Once model requirements are defined, organizations must ensure that the fulfillment of these requirements is documented appropriately throughout model development. However, requirements documentation remains a challenge for most organizations \cite{nahar2022collaboration}. Software engineers or system owners must express ML requirements, like a model accuracy or robustness goal, to model developers. In turn, model developers must be capable of providing evidence that these requirements are satisfied. While requirements documentation methods from software engineering are mature, these approaches must be modified to be appropriate for ML systems \cite{paleyes2022}. Examples of proposals for ML documentation include \textit{datasheets for datasets} \cite{gebru2021datasheets}, \textit{data version control} \cite{kuprieiev2021online}, and \textit{model cards} \cite{mitchell2019model}. However, these proposals focus on models and datasets rather than the requirements they must fulfill, failing to offer teams a documentation method that enforces requirements verification while maintaining comprehensibility.

\subsection{Problem 3: Implementing an Evidence-Based Evaluation}
\label{sec:infrastructure}

Once requirements are defined and the method by which they are documented is determined, software engineers and model developers must evaluate the model against these requirements and record evaluation results. On one side of this process, software engineers require an expressive way to encode requirements specific to the system under development. On the other side, model developers must translate requirements into measurable ML metrics and capture these measurements as evidence that requirements are satisfied. While most ML libraries include facilities for model evaluation \cite{scikit-learn}, appropriately implementing requirements beyond standard off\-line accuracy measures on held-out data still entails significant manual effort. A survey of existing research reveals that there is no ML-specific method for specifying and implementing system requirements \cite{villamizar2021requirements}, but examples exist. Breck et al. propose a rubric for quantifying the production readiness of ML systems that includes prescriptions for implementation far beyond just evaluating offline model accuracy \cite{breck2017ml}; the \textit{deepchecks} project offers a low-level approach to model requirements implementation by providing a library of \textit{checks} that developers can use to evaluate predictive performance~\cite{deepchecks}. Both of these tools provide useful functionality but fail to fully meet teams' needs because they do not associate the evidence they produce with requirements. Furthermore, they fail to couple a concrete implementation with flexibility sufficient to encompass any ML project. Standardizing and automating model evaluation through test infrastructure is important; organizations benefit from a process and tool that allows them to use selected metrics for model evaluation and export these results as evidence without disruption to the model development process.

\begin{figure*}[htbp]
\centering
\includegraphics[width=0.9\textwidth]{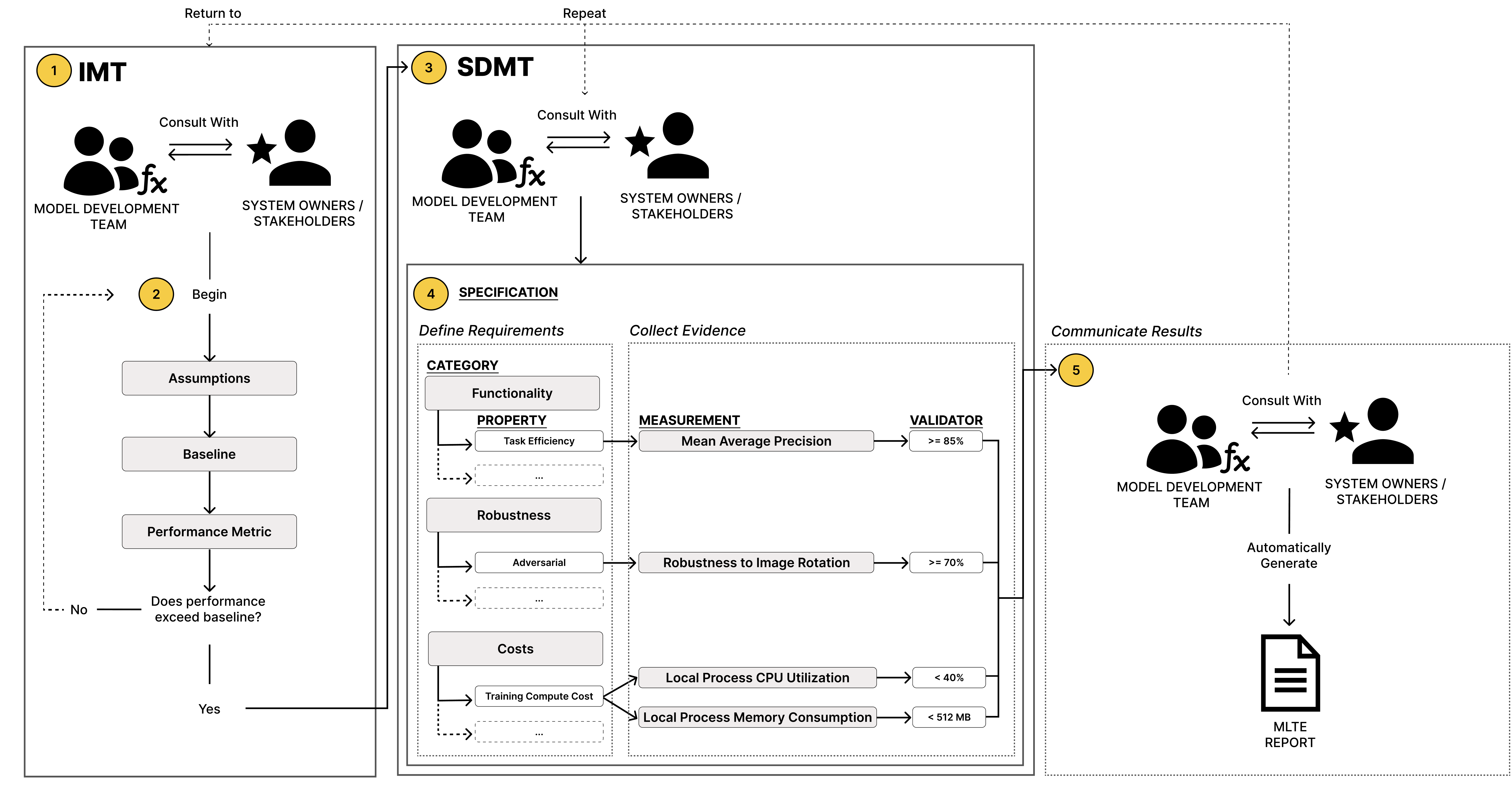}
\caption{The MLTE Workflow}
\label{mlte_fig}
\end{figure*}

In summary, previous work targeting these open problems is generally narrow in scope. Most attempts either design processes for particular aspects of ML model development or introduce a tool addressing specific elements, such as documenting a model's performance under certain data conditions. Our proposal differs by defining a repeatable and measurable process to evaluate ML systems that addresses the entire ML development life cycle.

\section{The MLTE Vision}

To address the three identified open problems in evaluating ML systems, we propose \textit{Machine Learning Test and Evaluation} (MLTE, pronounced ``melt''). MLTE consists of (a)~a framework through which organizations evaluate ML models in the context of their intended use and (b)~tooling that supports the implementation of this framework. MLTE is unique in that it gives organizations a \textit{process} from start to finish that leverages existing state-of-the-art research on testing ML system capabilities as well as interfaces for tracking experiments. MLTE is designed for \textit{interdisciplinary cross-team coordination} and encourages communication by offering specific \textit{collaboration points} and creating \textit{shared artifacts} (``method standardization'' and ``boundary objects''~\cite{star1989institutional}) throughout the ML development life cycle. 

In Figure~\ref{mlte_fig}, we outline the MLTE workflow. It starts with \sA~a negotiation between model developers and the model's consumers (system owners, other stakeholders) regarding model quality requirements; then \sB~goes through an iterative process of evaluating the model with regard to those requirements until performance exceeds an agreed-upon baseline. Next, \sC~model developers and all model and system stakeholders conduct an additional negotiation about model requirements beyond accuracy (e.g., robustness, fairness, inference costs); before \sD~evaluating those requirements; finally, \sE~teams use automated reporting functionality to facilitate communication of their ML evaluation results. 

We support each of the steps in the MLTE process with tooling. This tooling is implemented by embedding a domain-specific language for requirements specification and evidence collection in the Python programming language \cite{python_lang} and exposing this functionality via a library. The Python library allows practitioners to define, persist, and share requirements specifications. Furthermore, it provides functionality to generate, persist, and collect artifacts from ML procedures that can be used to verify that requirements are satisfied. Finally, it gives practitioners a mechanism to aggregate this evidence to generate a human-readable report. The MLTE process combined with this tooling provides an adaptive, open infrastructure to guide requirement generation, simplify testing, establish consistency, and communicate ML evaluation results. 

\section{Initial MLTE Prototype}

To meet its goal of facilitating ML system evaluation, the MLTE prototype consists of preliminary work on both the process and tooling. As described above, the MLTE process contains two primary phases supported by three interspersed collaboration points. In the following section, we detail each of these phases, the manner in which they address our open problems, and how we support each with  tooling.\footnote{The MLTE Package is available at \url{https://pypi.org/project/mlte-python}, the repository is available at \url{https://github.com/mlte-team/mlte}, and the framework is available at \url{https://github.com/mlte-team/mlte-framework}.} MLTE is still in its early stages, and as such we have not yet completed an initial empirical evaluation; our plans to do so are discussed in Section \ref{sec:future}.

MLTE is built with consideration to related and existing research on ML processes, such as typical ML workflows~\cite{amershi2019software}, the ML life cycle~\cite{ashmore2021assuring}, CRISP-DM~\cite{wirth2000crisp}, and TDSP~\cite{severtson2017}. These processes highlight best practices for developing ML systems, examine challenges inherent in bringing a ML system to production, and offer a template for teams to follow as they develop ML products and communicate with their stakeholders. All of these workflows mention requirements as a necessary input for ML development, and all include evaluation as a step in their ML process. Where MLTE diverges from this existing research is in its emphasis on a holistic framework to ensure effective evaluation of ML systems in the context of wider stakeholder and system goals. With its built-in communication points, emphasis on requirements definition, and integration with programmatic tooling, MLTE gives teams not only a rigorous process but also the tools needed to execute it.

\subsection{Internal Model Testing (IMT)}

At the first collaboration point (\sA\ in Fig.~\ref{mlte_fig}), the model development team and their customers identify model accuracy goals and suitable evaluation metrics as the first step of \textit{Internal Model Testing (IMT)}. The MLTE implementation is structured such that requirements definition directly informs how the model is subsequently evaluated. After completing initial collaboration, teams then proceed to the evaluation portion of IMT (\sB\ in Fig.~\ref{mlte_fig}). 

IMT gives teams an opportunity to refine their ML system implementation plan by conducting team-internal model evaluation. In this step, the MLTE approach balances prescriptiveness with flexibility by providing examples of common evaluations for different types of ML problems, which gives each team sufficient leeway to focus on the relevant concerns of their project. Teams select model characteristics from a curated collection that is organized by ML discipline (e.g., computer vision, natural language processing), and subsequently select a suitable method to assess their chosen characteristics. MLTE prompts teams to evaluate their model with their chosen baseline and performance metric(s), and MLTE tooling supports documentation of this process by providing the functionality to produce and record these results. IMT addresses Problem 2 (Documentation, see Sec.~\ref{sec:documentation}) and Problem 3 (Evaluation, see Sec.~\ref{sec:infrastructure}) by providing teams with a way to demonstrate that their model meets its requirements, and addresses Problem 1 (Communication, see Sec.~\ref{sec:comms-barriers}) through the collaboration points at \sA\ and~\sC. An initial exploratory phase in which performance goals evolve is common in ML projects~\cite{passi2018trust}; accordingly, MLTE provides two negotiation points so that teams can make an initial plan for requirements, and then update that plan and add requirements with respect to the system at collaboration point~\sC. Furthermore, we design IMT to cope with inherent uncertainty~\cite{breck2017ml, ozkaya2020really} by making the process iterative. We envision that teams make changes to their model and execute the IMT process as frequently as needed until the model is ready for \textit{System Dependent Model Testing}.

\subsection{System Dependent Model Testing (SDMT)}
\label{sec:SDMT}

The second phase of MLTE, \textit{System Dependent Model Testing (SDMT)} (\sC\ and \sD\ in Fig.~\ref{mlte_fig}), provides a structure and process for evaluating model capabilities and limitations in the context of the \textit{system into which it will be integrated}. The process starts with a collaboration point (\sC\ in Fig.~\ref{mlte_fig}), during which model development teams, stakeholders, and system owners describe requirements by defining a \textit{specification}. Teams construct a specification by selecting the characteristics the model must exhibit to be considered acceptable. These characteristics, which we refer to as \textit{properties}, may be any attribute of the trained model, the procedure used to train it (including training data), or its ability to perform inference. A property is `abstract' in the sense that there may be many ways in which it might be assessed. To assist teams in defining their requirements, the MLTE framework currently organizes properties into three categories: functionality, robustness, and costs. These categories will expand to include new techniques and categories as research advances. The applicability of the properties within a given category depends on the type of ML task, the objective of the system into which the model is integrated, stakeholder priorities, and value judgments \cite{raji2020closing}. This process, which occurs at collaboration point \sC, addresses Problem 1 (Communication) by establishing a communication waypoint during which system dependent model requirements must be negotiated. Once populated with properties, the specification itself serves as an artifact that addresses Problem 2 (Documentation).

MLTE tooling supports requirements negotiation by providing the ability to programmatically define and persist specifications. Listing \ref{fig:specification} shows use of the MLTE library to perform this task. Users import types that represent individual properties and combine instances of these types within a \texttt{Spec} object to assemble a new specification. Completed \texttt{Spec} objects may then be persisted to an external \textit{artifact store} and subsequently restored for further inspection and manipulation. The ability to programmatically define a specification addresses Problem 3 (Evaluation) by standardizing the process by which it is generated and defining its interface.

\begin{listing}[t]
\begin{minted}[frame=lines,tabsize=2,breaklines,fontsize=\footnotesize]{python}
import mlte
from mlte.specification import Spec
from mlte.property import (
    TaskEfficacy,
    TrainingComputeCost
)
# Initialize the session
mlte.set_model("ModelName", "v0.0.1")
mlte.set_artifact_store_uri("localhost:8080")
# Define the specification
spec = Spec(
    TaskEfficacy(),
    TrainingComputeCost()
)
# Persist the specification to artifact store
spec.save()
\end{minted}
\caption{Defining a \textit{specification} with MLTE.}
\label{fig:specification}
\end{listing}

Following requirements definition, MLTE prescribes an evidence-based approach to requirements satisfaction (\sD\ in Fig.~\ref{mlte_fig}). In this paradigm, model developers determine how the model performs against the requirements set by system owners by collecting \textit{results} that attest to the satisfaction of particular properties. Practitioners produce results by defining and executing \textit{measurements} --- functions that assess phenomena related to the property of interest. Model developers define the range of acceptable values for a particular result by defining one or more associated \textit{validators}. A validator is a function that accepts a measurement's output, compares the observed output with an acceptable value or range of values, and produces a result that reflects this comparison. Together, properties, measurements, results, and validators ensure that teams generate evidence to inform larger decisions about the functionality of a system. 

This evidence-based approach to requirements satisfaction is labor-intensive, so our programmatic tooling focuses much of its functionality on easing the burden of implementation. To simplify the evaluation process, we provide a small collection of measurement definitions that might not be offered by existing libraries. More importantly, the tooling implements an infrastructure for capturing results produced by any ML library and persisting them in a stable, machine-readable format. Practitioners may utilize existing adaptors defined by the MLTE library to achieve this integration, or they may develop their own to add support for user-defined measurements. Listing \ref{fig:evidence} demonstrates both aspects of this functionality. Users import internal measurements and evaluate these to produce results, or wrap the output of external measurements in a suitable \texttt{Result} type. These \texttt{Result} instances are then persisted to the same artifact store that houses completed specifications.

\begin{listing}[t]
\begin{minted}[frame=lines,tabsize=2,breaklines,fontsize=\footnotesize]{python}
# Integrate a result from an external library
from sklearn.metrics import accuracy_score
from mlte.measurement.result import Real
accuracy = Real("accuracy", accuracy_score(...))
# Utilize MLTE-provided measurement
from mlte.measurement.cpu import ProcessCPUUtilization
m = ProcessCPUUtilization("cpu stats")
cpu_stats = m.evaluate(start_training_job())
# Persist results to artifact store
Result.save_all(accuracy, cpu_stats)
\end{minted}
\caption{Generating evidence with internal and external measurements.}
\label{fig:evidence}
\end{listing}

The MLTE process concludes with the generation of a \textit{report} that encapsulates all of the knowledge gained about the model and the system as a consequence of the evaluation process (\sE\ in Fig.~\ref{mlte_fig}). Report production, demonstrated in Listing \ref{fig:report}, ensures that teams have a method through which they can both examine and display the results of their work. Users load previously-saved results from the artifact store, validate these results with provided methods, and combine the output from validation with a specification to create a report. Reports give teams a shared artifact to understand the model, its capabilities, and its context, assisting downstream integration and maintenance activities. This allows teams to avoid barriers to creating shared understanding of results described in Problem 1 (Communication) by giving them a method and infrastructure for communicating their results. 

We design the SDMT section of MLTE to support diverse ML projects, regardless of discipline. All elements of the MLTE hierarchy (properties, measurements, results, validators) are collections of functions that can be expanded by MLTE users. Such expansions may be maintained internally (e.g., a team-internal repository of custom measurements) or may be contributed back to the MLTE ecosystem by updating the package itself. Because extensibility is a primary feature, MLTE allows users to rapidly integrate new functionality to suit their specific needs. We envision our existing MLTE tooling as the beginnings of an infrastructure for ML evaluation, rather than as a fully-realized toolkit.

\begin{listing}[t]
\begin{minted}[frame=lines,tabsize=2,breaklines,fontsize=\footnotesize]{python}
# Load the specification
from mlte.specification import Spec
spec = Spec.load()
# Load result(s)
from mlte.measurement.result import Real
accuracy = Real.load("accuracy")
# Validate results, generate report
report = spec.bind(
    {"TaskEfficacy": ["accuracy"]}
    accuracy.greater_than(0.85)
)
report.save()
\end{minted}
\caption{Combining evidence in a \textit{report}.}
\label{fig:report}
\end{listing}

\section{Future Plans}
\label{sec:future}

Overall, the MLTE framework provides guidance that enforces specific process steps while retaining flexibility. Future research will interrogate this tension between rigor and flexibility. In particular, we plan to study the role of shared machine-readable requirements specifications and reported results as boundary objects~\cite{star1989institutional} between teams. Additionally, MLTE is still in an early stage of usage; the evaluation framework and supporting tooling still need to be evaluated by practitioners. To this end, we will conduct an interview study that addresses two inquiries: how organizations are currently evaluating their ML systems, and how MLTE would function in the context of their organizational processes. Input from this study will inform the refinement of both the framework and tooling as we seek to maximize versatility and capability while maintaining ease-of-use.

\section*{Acknowledgements}

This work was supported in part by the U.S. Army Combat Capabilities Development Command (DEVCOM) Army Research Laboratory under Support Agreement No. USMA21050, the U.S. Army DEVCOM C5ISR Center under Support Agreement No. USMA21056, and the U.S. Air Force Research Laboratory under Support Agreement No. USMA2226. The views expressed in this paper are those of the authors and do not reflect the official policy or position of the United States Military Academy, United States Army, United States Department of Defense, National Science Foundation, or United States Government. 

Lewis' work was funded and supported by the Department of Defense under Contract No. FA8702-15-D-0002 with Carnegie Mellon University for the operation of the Software Engineering Institute, a federally funded research and development center (DM23-0017). Kästner's work is supported in part by the National Science Foundation (awards 1813598, 2131477, and 2206859). 

\bibliographystyle{IEEEtran}
\bibliography{main}

\end{document}